\documentclass[12pt]{emulateapj}

\usepackage{graphicx,amsmath,amsfonts,amssymb,subfigure,hyperref}

\begin{document}

\title{Exploring the chemical link between local ellipticals and their high-redshift progenitors}

\slugcomment{Accepted to ApJ Letters}
\author{Joel Leja\altaffilmark{1}, Pieter G. van Dokkum\altaffilmark{1}, Ivelina Momcheva\altaffilmark{1}, Gabriel Brammer\altaffilmark{2}, Rosalind E. Skelton\altaffilmark{3}, Katherine E. Whitaker\altaffilmark{4}, Brett H. Andrews\altaffilmark{5}, Marijn Franx\altaffilmark{6}, Mariska Kriek\altaffilmark{7}, Arjen van der Wel\altaffilmark{8}, Rachel Bezanson\altaffilmark{9}, Charlie Conroy\altaffilmark{10}, Natascha F\"{o}rster Schreiber\altaffilmark{11}, Erica Nelson\altaffilmark{1}, Shannon G. Patel\altaffilmark{6}}

\altaffiltext{1}{Department of Astronomy, Yale University, New Haven, CT 06511, USA} 
\altaffiltext{2}{European Southern Observatory, Alonson de C\'{o}rdova 3107, Casilla 19001, Vitacura, Santiago, Chile}
\altaffiltext{3}{South African Astronomical Observatory, Observatory Road, Cape Town, South Africa}
\altaffiltext{4}{Astrophysics Science Division, Goddard Space Center, Greenbelt, MD 20771, USA}
\altaffiltext{5}{Department of Astronomy, The Ohio State University, Columbus, OH 43210, USA}
\altaffiltext{6}{Leiden Observatory, Leiden University, Leiden, Netherlands}
\altaffiltext{7}{Astronomy Department, University of California at Berkeley, Berkeley, CA 94720}
\altaffiltext{8}{Max-Planck Institute for Astronomy (MPIA), K\"{o}nigstuhl 17, D-69117 Heidelberg, Germany}
\altaffiltext{9}{Steward Observatory, University of Arizona, 933 North Cherry Avenue, Tucson, AZ 85721, USA}
\altaffiltext{10}{Department of Astronomy \& Astrophysics, University of California, Santa Cruz, CA, USA}
\altaffiltext{11}{Max-Planck-Institut f\"{u}r extraterrestrische Physik, Giessenbachstrasse, D-85748 Garching, Germany}
\begin{abstract}
We present Keck/MOSFIRE K-band spectroscopy of the first mass-selected sample of galaxies at $z\sim2.3$. Targets are selected from the 3D-HST Treasury survey. The six detected galaxies have a mean [NII]$\lambda$6584/H$\alpha$ ratio of $0.27\pm0.01$, with a small standard deviation of 0.05. This mean value is similar to that of UV-selected galaxies of the same mass. The mean gas-phase oxygen abundance inferred from the [NII]/H$\alpha$ ratios depends on the calibration method, and ranges from 12+log(O/H)$_{\mathrm{gas}}$$=8.57$ for the {Pettini} \& {Pagel} (2004) calibration to 12+log(O/H)$_{\mathrm{gas}}$$= 8.87$ for the {Maiolino} {et~al.} (2008) calibration. Measurements of the stellar oxygen abundance in nearby quiescent galaxies with the same number density indicate 12+log(O/H)$_{\mathrm{stars}}$$= 8.95$, similar to the gas-phase abundances of the $z\sim2.3$ galaxies if the {Maiolino} {et~al.} (2008) calibration is used. This suggests that these high-redshift star forming galaxies may be progenitors of today's massive early-type galaxies. The main uncertainties are the absolute calibration of the gas-phase oxygen abundance and the incompleteness of the $z\sim2.3$ sample: the galaxies with detected H$\alpha$ tend to be larger and have higher star formation rates than the galaxies without detected H$\alpha$, and we may still be missing the most dust-obscured progenitors.
\end{abstract}
\keywords{
galaxies: abundances ---
galaxies: evolution
}
\section{Introduction}

Observations of the elemental abundances of galaxies provide information on the build-up of metals in the Universe and on the importance of winds and feedback ({Dav{\'e}}, {Finlator}, \&  {Oppenheimer} 2012). Most studies find that the mass-metallicity relation evolves with redshift, such that at fixed stellar mass, galaxies have lower metallicity at earlier times ({Erb} {et~al.} 2006b; {Maiolino} {et~al.} 2008; {Zahid} {et~al.} 2013, though see also {Stott} {et~al.} 2013). This is consistent with expectations from simple models in which gas is gradually enriched by (post-)AGB stars and supernovae.

In addition to measuring their gas-phase metallicities, it is also possible to measure the {\em stellar} metallicities of galaxies ({Gallazzi} {et~al.} 2005, {Panter} {et~al.} 2008, {Conroy}, {Graves}, \& {van  Dokkum} 2013 and references therein). As the stellar metallicities reflects the gas-phase metallicities at the time of star formation, the combined measurements of stellar and gas-phase metallicities over cosmic time puts powerful constraints on galaxy formation models ({Bresolin} {et~al.} 2009; {Sommariva} {et~al.} 2012).

In this Letter we take a step in this direction by comparing the stellar oxygen abundances of massive galaxies in the local Universe to the gas-phase oxygen abundances of their putative progenitors at early times. This comparison should be relatively straightforward for massive galaxies, as they formed most of their stars at redshifts $z\gtrsim 2$ ({Thomas} {et~al.} 2005; {Conroy} {et~al.} 2013). Therefore, there should be a direct correspondence between the gas-phase metallicities of massive galaxies at $z\gtrsim 2$ and the stellar metallicities of their descendants at $z=0$.

This project has recently become possible due to the CANDELS and 3D-HST datasets, which provide mass-limited samples with accurate redshifts, and to the advent of the MOSFIRE spectrograph on the Keck telescope ({McLean} {et~al.} 2012). Furthermore, accurate stellar abundances of individual elements have recently been derived from averaged spectra of SDSS galaxies of different masses ({Conroy} {et~al.} 2013).

In order to link progenitor galaxies with their descendants we require that they have the same cumulative number density ({van Dokkum} {et~al.} 2010; {Patel} {et~al.} 2013; {Leja}, {van Dokkum}, \& {Franx} 2013). This comparison at a constant number density is preferable to comparison at constant stellar mass, as it explicitly takes the mass evolution of galaxies into account.

Throughout the paper, we assume a Chabrier IMF ({Chabrier} 2003) and a $\Lambda$CDM cosmology with $H_0 = 70$ km s$^{-1}$ Mpc$^{-1}$, $\Omega_M$ = 0.3, and $\Omega_{\Lambda}$ = 0.7.

\section{Sample Selection}

\begin{figure*}[t]
\begin{center}
\includegraphics[scale=0.8, bb=0 270 600 510]{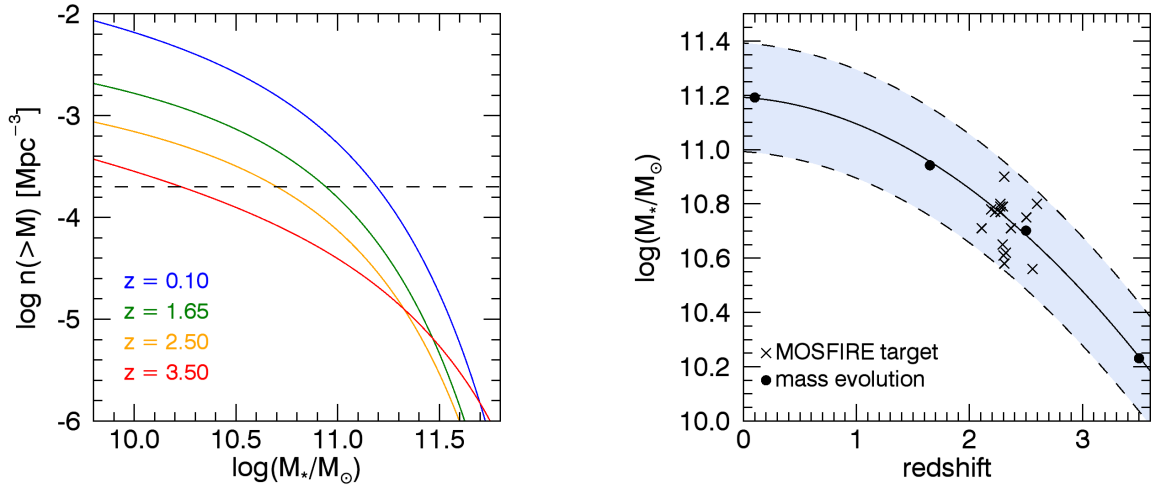}
\caption{Stellar mass evolution is inferred at a constant cumulative number density of $2\times10^{-4}$ Mpc$^{-3}$ (dashed horizontal line) from the mass functions of {Marchesini} {et~al.} (2009) (colored lines). The right panel shows the mass evolution at this number density (solid circles) and a quadratic parameterization (solid line, Equation 1). The selection area ($\pm$0.2 dex) is shaded in light blue, and target galaxies are marked with crosses.}
\label{nds}
\end{center}
\end{figure*}

We select spectroscopic targets in the UKIDSS-UDS field from version 2.1 of the 3D-HST survey catalogs ({Brammer} {et~al.} 2012). The 3D-HST catalogs contain redshifts and stellar masses derived from a combination of HST/G141 grism spectra and deep photometric data, with wavelength coverage from UV to Spitzer/IRAC as described in Skelton et al.\ (in prep). Reported stellar masses are the current mass in stars and stellar remnants. When possible, the star formation rates (SFRs) are based on the UV+IR flux, with the IR determined from Spitzer/MIPS; otherwise, for fainter objects, they come from fits of stellar population synthesis models to the stellar spectral energy distributions (SEDs). As shown in {Wuyts} {et~al.} (2011) the techniques give consistent answers where they overlap. We note that H$\alpha$ emission in detected galaxies constitutes a median of 5\% of the $K$-band flux, and thus has a negligible effect on broadband photometry. Catalog information and emission line properties for the targeted sample are shown in Table 1.

The targets are selected at a fixed cumulative number density of $2 \times 10^{-4}$ Mpc$^{-3}$ in the stellar mass functions of {Marchesini} {et~al.} (2009), which are $\sim$100\% complete in the relevant mass and redshift range. Selecting galaxies at a constant number density will effectively link galaxies across different redshifts if the stellar mass rank order of galaxies is approximately conserved with time, or alternatively, if processes that break rank order (i.e. merging or scatter in growth rates) have little effect on average galaxy properties. {Leja} {et~al.} (2013) found this technique is effective in predicting the median descendant stellar mass from $z=3$ to $z=0$ to $\le0.15$ dex in a semi-analytical model. This selection corresponds to a redshift-dependent stellar mass criterion, shown in Figure \ref{nds}. The stellar mass evolution is parametrized as:
\begin{equation}
\log{\left(\frac{M_*(z)}{M_{\odot}}\right)} = 11.19 - 0.03z - 0.07z^2
\end{equation}
Specifically, we target galaxies that have stellar masses within $\pm$0.2 dex of this relationship. We further require that H$\alpha$ and [NII] fall both within the $K$-band filter and on the MOSFIRE detector, effectively creating a joint constraint on the redshift and the ($\alpha$, $\delta$) of each target. These selection criteria result in a mask with 16 galaxies, with redshifts 2.1 $< z <$ 2.55.

\begin{center}
\begin{deluxetable*}{cccccccccccc}[b!h]
\tablecaption{Spectroscopic Targets}
\tablenum{1}
\tablehead{\colhead{ID} & \colhead{$\alpha$} & \colhead{$\delta$} & \colhead{F140W} & \colhead{$\mathcal{R}^a$} & \colhead{log(M$_*$)} & \colhead{SFR} & \colhead{r$_e$$^b$} & \colhead{z$_{\mathrm{3DHST}}$} & \colhead{z$_{\mathrm{spec}}$} & \colhead{[NII]/H$\alpha$} \\ 
\colhead{ } & \colhead{(J2000)} & \colhead{(J2000)} & \colhead{} & \colhead{} & \colhead{(M$_{\odot}$)} & \colhead{(M$_{\odot}$/yr)} & \colhead{(kpc)} & & & }
\startdata
2522 & 34.42735 & -5.26471 & 22.69 & 24.34 & 10.61 & 242 & 5.26 & 2.30 & 2.315 & 0.29$^{+0.03}_{-0.02}$ \\
5698 & 34.44652 & -5.24892 & 23.55 & 26.24 & 10.71 & 141 & 2.82 & 2.11 & 2.127 & 0.31$^{+0.03}_{-0.03}$ \\
8461 & 34.42749 & -5.23648 & 23.66 & 26.24 & 10.78 & 142 & 3.31 & 2.19 & 2.299 & 0.28$^{+0.03}_{-0.02}$ \\
10746 & 34.39999 & -5.22679 & 22.39 & 24.09 & 10.71 & 88. & 6.31 & 2.37 & 2.541 & 0.16$^{+0.03}_{-0.03}$ \\
19440 & 34.41029 & -5.18816 & 23.27 & 25.17 & 10.77 & 30. & 4.63 & 2.28 & 2.291 & 0.28$^{+0.02}_{-0.02}$ \\
24828 & 34.42530 & -5.16451 & 23.30 & 24.93 & 10.80 & 67. & 5.90 & 2.27 & 2.243 & 0.31$^{+0.03}_{-0.03}$ \\
3956 & 34.46006 & -5.25758 & 23.50 & 26.57 & 10.79 & 0 & 1.46 & 2.27 & -- & -- \\
5326 & 34.43100 & -5.25042 & 26.08 & 27.54 & 10.56 & 1 & 3.19 & 2.56 & -- & -- \\
9277 & 34.42780 & -5.23286 & 24.41 & 26.77 & 10.80 & 0 & 1.61 & 2.60 & -- & -- \\
10771 & 34.42257 & -5.22659 & 23.55 & 25.21 & 10.65 & 21 & 3.73 & 2.29 & -- & -- \\
11700 & 34.44719 & -5.22188 & 24.57 & 26.82 & 10.77 & 9 & 2.71 & 2.23 & -- & -- \\
11909 & 34.42215 & -5.22073 & 23.88 & 27.18 & 10.75 & 0 & 0.63 & 2.50 & -- & -- \\
12447 & 34.40580 & -5.21886 & 22.79 & 25.95 & 10.79 & 1 & 0.86 & 2.30 & -- & -- \\
16478 & 34.40424 & -5.20082 & 24.33 & 27.35 & 10.62 & 0 & 3.97 & 2.32 & -- & -- \\
18367 & 34.38511 & -5.19237 & 23.60 & 27.07 & 10.58 & 0 & 0.76 & 2.31 & -- & -- \\
22984 & 34.41577 & -5.17188 & 23.81 & 26.42 & 10.90 & 269 & 2.44 & 2.31 & -- & -- \\
\enddata
\tablecomments{$^a$ $\mathcal{R}$-band magnitude, defined in {Steidel} \& {Hamilton} (1992)\\
$^b$ Sizes calculated in the H-band as described in {van der Wel} {et~al.} (2012)}
\end{deluxetable*}
\end{center}
\section{Observations and data reduction}
The MOSFIRE K-band observations were conducted on January 20th, 2013, with $\sim$1.5" seeing. An ABAB dither pattern with a 1.5" nod was used. Slit widths were 0.7". A single mask was observed with 16 targets for 85 minutes, with 6 of the targets showing clear line emission. We estimate that sky lines obscure only $\sim$3\% of the spectral range for emission lines with a central per-pixel S/N=20, typical for detected emission lines. The non-detections are thus likely caused by intrinsically weak or dust-obscured galaxy emission lines, rather than overlap of intrinsically bright lines with sky emission features.

The MOSFIRE data reduction pipeline\footnote{\url{https://code.google.com/p/mosfire/}} was used to reduce the spectroscopic data. The pipeline performs flat fielding, wavelength calibration, sky subtraction, and cosmic ray removal before producing a final two-dimensional output with an associated variance map. One dimensional spectra were extracted using the optimal extraction method of {Horne} (1986). No flux calibration or reddening correction was necessary for this study.

\begin{figure*}
\begin{center}
\includegraphics[scale=0.8, bb=0 150 600 620]{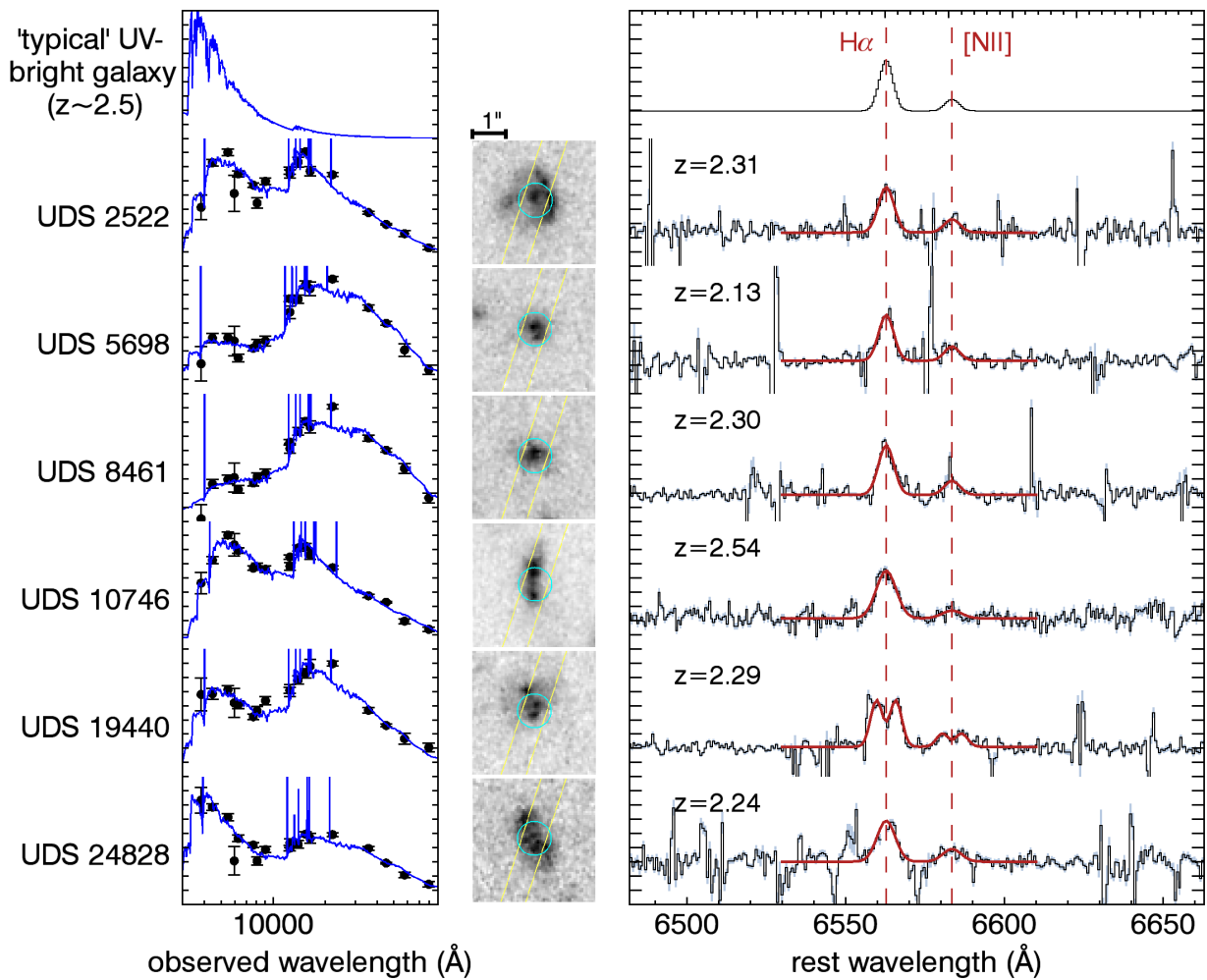}
\caption{Template SEDs (blue) fit to the broadband photometry (black) are shown for each galaxy with detected emission lines. The units are arbitrarily normalized in f$_{\lambda}$. The center column has HST/F160W cutouts with the MOSFIRE slit orientation overlaid. The MOSFIRE K-band spectra are on the right, with 1$\sigma$ errors in blue. Single Gaussian fits to the H$\alpha$ and [NII] line profiles are overlaid in red. UDS-19440 is fit with a double Gaussian to account for the rotational line profile. We also include the SED of a typical UV-bright galaxy, constructed with the stellar population synthesis models of {Bruzual} \& {Charlot} (2003) and typical UV-bright stellar population parameters from {Papovich}, {Dickinson}, \&  {Ferguson} (2001).}
\label{spectra}
\end{center}
\end{figure*}

The K-band spectra for targets with detected emission lines are shown in Figure \ref{spectra}, along with spectral energy distributions (SEDs) and F160W direct images. H$\alpha$ and [NII] emission lines are fit with Gaussian profiles; the only exception is UDS-19440, which is fit with a double Gaussian to properly model the line profile. [NII] and H$\alpha$ are fit simultaneously, with their line widths and redshifts constrained to the same value to improve accuracy when fitting the weaker [NII] line. The adopted fluxes are the areas of the Gaussians. Errors in the line profile are determined by perturbing each flux value within a Gaussian probability distribution, then remeasuring the line profile. The width of the Gaussian probability distribution is set to the $1\sigma$ flux error in the pixel. The errors on measured parameters are taken as the 68\% range in derived parameters over 1000 iterations of perturbed spectra.

\section{Results}
\subsection{[NII]/H$\alpha$ ratios}
We measure a mean [NII]/H$\alpha$ ratio of 0.27 $\pm$ 0.01 (error in the mean) $\pm$ 0.05 (standard deviation) in our sample.

\begin{figure*}[t!]
\begin{center}
\includegraphics[scale=0.8, bb=0 0 600 800]{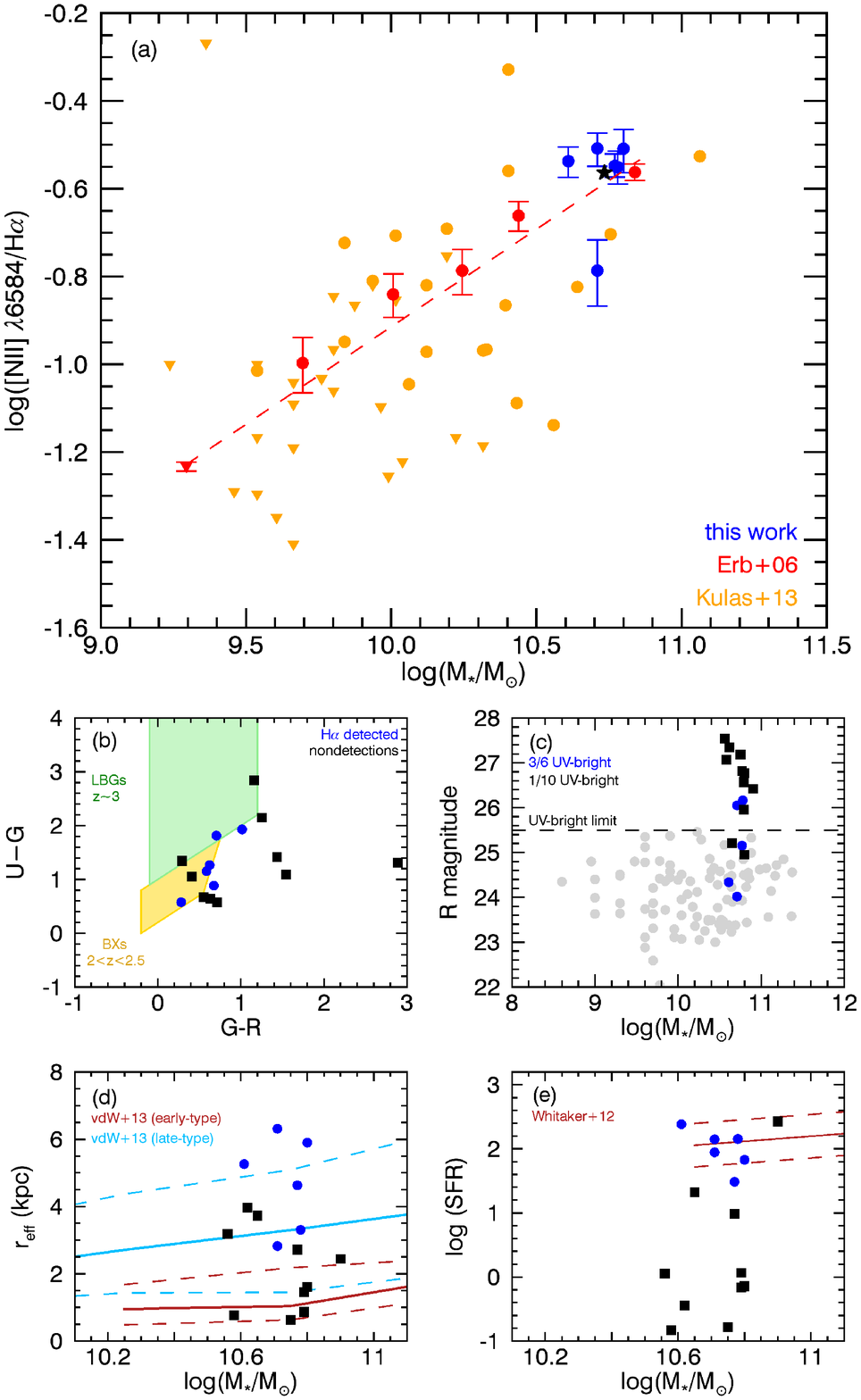}
\caption{In all panels, blue circles are galaxies with detected emission lines, while black squares are non-detections. (a) Comparing [NII]/H$\alpha$ ratios from our mass-selected sample (blue) to those from UV-selected samples (orange, red). The mean of our mass-selected sample is marked with a black star. Galaxies with upper limits on [NII]/H$\alpha$ are marked with downward triangles. The {Erb} {et~al.} (2006a) data are from composite spectra. The red dashed line is a linear fit to the {Erb} {et~al.} (2006a) data. (b) The UV-bright color-color selection boxes are taken from {Steidel} {et~al.} (2004). (c) Grey points are from the UV-bright sample of {Erb} {et~al.} (2006b), and the UV-bright magnitude limit is indicated with a dashed line. (d) The size-mass relations are taken from S\'{e}rsic fits to the circularized H-band light profile at $z\sim2.25$ (van der Wel, in prep): median values are shown as solid lines, while the 16th and 84th percentiles are dashed lines. (e) The starforming sequence is taken from {Whitaker} {et~al.} (2012) and marked with a solid red line, with 1$\sigma$ scatter denoted by dashed lines.}
\label{ratios}
\end{center}
\end{figure*}

We compare the relationship between stellar mass and [NII]/H$\alpha$ in our sample and in UV-selected samples (Figure \ref{ratios}). By contrasting data rather than derived quantities, we cleanly assess potential differences in [NII]/H$\alpha$ ratio between different samples.

Specifically, we compare our measurements to those of {Erb} {et~al.} (2006a) and {Kulas} {et~al.} (2013). These studies select spectroscopic targets to be bright in the rest-frame UV ($\mathcal{R}$ $<$ 25.5) and to fulfill color-color criteria in the rest-frame UV, as described in {Steidel} {et~al.} (2004) and {Adelberger} {et~al.} (2004). The data from {Erb} {et~al.} (2006a) are stacked spectra with $\sim$15 galaxies per point.

The stellar masses in {Erb} {et~al.} (2006a) are reported as the integral of the SFR. In order to compare with our data, we convert the {Erb} {et~al.} (2006a) stellar masses into the mass in stars and stellar remnants using the following formula:
\begin{equation}
\log{\left( \frac{M}{M_{Erb}} \right) } = 1.06 - 0.24T + 0.01 T^2
\end{equation}
with $T \equiv \log{\left(\mathrm{age/yr}\right)}$, and ages taken from {Erb} {et~al.} (2006a). The formula is a fit to the mass loss rates in the {Bruzual} \& {Charlot} (2003) models for a Chabrier IMF, and applies only to ages $>$ 2 Myr.

Figure \ref{ratios} shows that mass-selected galaxies have similar [NII]/H$\alpha$ ratios compared to UV-selected galaxies. We quantify the significance of this result by simulating our observations using population statistics from the UV-selected samples. First, we fit a linear relationship to the {Erb} {et~al.} (2006a) points, finding:
\begin{equation}
\log{\mathrm{([NII]/H}\alpha)} = -5.36 + 0.44\log{(\mathrm{M_{*}}/\mathrm{M}_{\odot})}
\end{equation}
We use this relationship to estimate the UV-selected [NII]/H$\alpha$ ratio at the average stellar mass of our sample: log($<$M$>$$_*$/M$_{\odot}$) = 10.73. We calculate the biweight scatter ({Beers}, {Flynn}, \& {Gebhardt} 1990) about the relationship to be 0.22 dex, using the {Kulas} {et~al.} (2013) galaxies with M$_{*} > 10^{10}$ M$_{\odot}$. We then simulate our observations by repeatedly sampling six galaxies from a Gaussian probability distribution with a mean and scatter fixed to the values above. This results in a mean [NII]/H$\alpha$ ratio greater than our measured ratio only 40\% of the time. This mass-selected sample thus does not have significantly higher [NII]/H$\alpha$ ratios than UV-selected samples.

To further explore whether mass-selected, H$\alpha$-detected galaxies are different from UV-selected galaxies of the same mass, we include two panels in Figure \ref{ratios} that explore what fraction of the mass-selected sample does not fulfill the selection criteria for a UV-bright sample. Our sample is split into galaxies with detected emission lines and non-detections. $\mathcal{U,G,}$ and $\mathcal{R}$ magnitudes are measured directly from the best-fit EAZY template ({Brammer}, {van Dokkum}, \&  {Coppi} 2008), then shown relative to the UV-bright selection criteria. The results indicate that 50\% of the galaxies with detected emission lines fit the UV-bright selection criteria\footnote{If we instead ask whether these galaxies would be considered UV-bright when placed at {\it any} redshift, this changes to 67\%.}, but only 10\% of galaxies without detected emission lines fit the same criteria. The whole sample is thus primarily UV-faint: however, 50\% of the galaxies with detected line emission are UV-bright.


\subsection{Oxygen abundances}
We next convert the measured [NII]/H$\alpha$ ratios into oxygen abundances.  To demonstrate the spread in oxygen abundance between metallicity calibrations, we calculate the oxygen abundance in three different calibration systems: {Maiolino} {et~al.} (2008) (M08), {Denicol{\'o}}, {Terlevich}, \&  {Terlevich} (2002) (D02), and {Pettini} \& {Pagel} (2004) (PP04).

The M08 relationship between observed [NII]/H$\alpha$ ratio and oxygen abundance is:
\begin{equation}
\log{(\mathrm{[NII]/H\alpha)}} = c_0 + c_1x +c_2x^2 +c_3x^3 +c_4x^4
\end{equation}
with $c_0 = -0.7732$, $c_1=1.2357$, $c_2=- 0.2811$, $c_3=-0.7201$, $c_4=-0.333$, and $x\equiv12+\log{\mathrm{(O/H)}}-8.69$. The scatter in this conversion is taken to be 0.1 dex.


We find a mean oxygen abundance of 12 + log(O/H) = $8.87 \pm 0.04$, 8.70$\pm 0.10$ and 8.57$\pm 0.08$ for the M08, D02, and PP04 calibrations respectively. The quoted error is the error in the mean, while the standard deviations are 0.09, 0.08, and 0.08, again respectively. These mean abundances range from 0.12 dex below the solar value of 12 + log(O/H) = 8.69 ({Asplund} {et~al.} 2009) for the PP04 calibration to 0.18 dex above solar for M08.

\subsection{Comparison to stellar abundances at $z=0$}
We now compare the gas-phase oxygen abundances to the stellar oxygen abundances of nearby galaxies. We adopt the stellar oxygen abundances measured in {Conroy} {et~al.} (2013). This study analyzes spectra from the inner 0.4-0.8 effective radii of local quiescent galaxies stacked in bins of stellar velocity dispersion, and fits a full-spectrum model to them, described in {Conroy} \& {van Dokkum} (2012). The model constrains the abundances of individual elements, including oxygen. To compare with our results, we derive the average stellar mass for these stacks, and interpolate the stellar oxygen abundance at the expected descendant mass. Specifically, we interpolate between the two bins with log($\sigma$/km/s) = 2.39 and 2.47, with corresponding stellar masses of 10.96 and 11.34. This results in an oxygen abundance of 12 + log(O/H) $= 8.95^{+0.03}_{-0.03}$. The error bars represent the stellar oxygen abundances as inferred at the edge of the mass selection box (see Figure \ref{nds}).

We can now compare these $z=0$ stellar abundances to the high redshift gas-phase metallicities derived in 4.2. If the high redshift galaxies are progenitors of the low redshift galaxies, and we are observing the main epoch of star formation in the high redshift galaxies, then the high redshift gas-phase abundance should match the low redshift stellar metallicity. Interestingly, the gas-phase metallicities are lower than the stellar abundances, with the difference depending on the calibration method. PP04 produces the largest inconsistency, with the stellar abundance nearly 0.4 dex higher than the gas metallicity. The best match comes from the M08 calibration, which produces gas metallicities that are only 0.08 dex lower than the stellar abundance.

\begin{figure}[t!]
\begin{center}
\includegraphics[scale=0.35,bb=0 100 600 700]{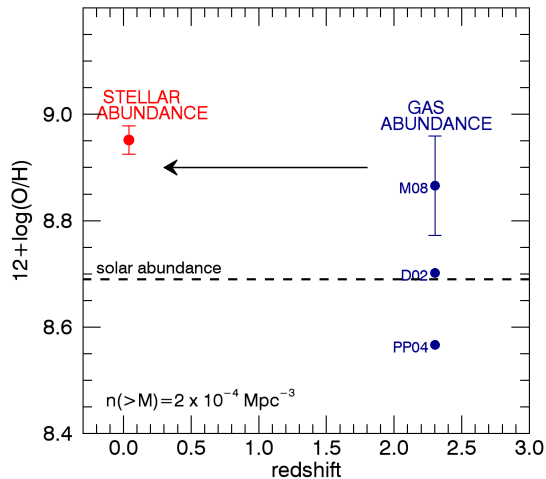}
\caption{Comparing the gas-phase oxygen abundances of galaxies at z $\sim2.3$ to the stellar abundances of local quiescent galaxies at the same number density. The mean gas-phase metallicities from multiple metallicity calibrations are shown in dark blue. The scatter in these calibrations is $\sim$0.1-0.2 dex. The solar abundance is from {Asplund} {et~al.} (2009). The gas-phase error bar is the standard deviation of the metallicities; it is only shown for the M08 calibration for clarity.}
\label{metals}
\end{center}
\end{figure}

\section{Discussion}
\label{disc}
In this Letter, we have measured [NII]/H$\alpha$ ratios from Keck/MOSFIRE K-band spectroscopy of a mass-selected sample at high redshift. We demonstrate that these [NII]/H$\alpha$ ratios are similar to those measured in surveys of UV-bright galaxies. We measure an average [NII]/H$\alpha$ ratio of 0.27, corresponding to an average gas-phase oxygen abundance of 12+log(O/H) = 8.57-8.87, depending on the metallicity calibration adopted. Taking Figure 4 at face value, we would conclude that the M08 calibration gives the best match to the $z=0$ stellar abundances, and should therefore be preferred over the other calibrations. However, there are many sources of systematic uncertainty in this study.

We first consider the sources of the differences in the three calibrations. The M08 oxygen abundance is calculated from a combination of the direct $T_e$ method for galaxies with log(O/H) + 12 $<$ 8.35, and with strong line ratios and the photoionization models of {Kewley} \& {Dopita} (2002) for log(O/H) + 12 $>$ 8.35. D02 derives oxygen abundances primarily from the direct $T_e$ method, with roughly one third of their abundances from oxygen or sulfur strong line ratios. PP04 uses 97\% direct $T_e$ abundances. All three studies then derive a relationship between [NII]/H$\alpha$ ratio and oxygen abundance.

The spread in metallicity calibrations may be related to issues with the photoionization models, or temperature gradients and fluctuations that cause metallicities determined by the direct $T_e$ method to be underestimated ({Peimbert} 1967; {Kewley} \& {Ellison} 2008). Observational studies of optical recombination lines show a systematic differences of $0.26\pm0.09$ dex with the $T_e$ method ({Esteban} {et~al.} 2009), which may explain some of the variation.

There is also emerging evidence that, at high redshift, metallicities based on [NII]/H$\alpha$ are offset from those based on [OII]$\lambda$3727, [OIII]$\lambda4959,5007$ and H$\beta$, even within the same system of metallicity calibrations ({Newman} {et~al.} 2013; {Cullen} {et~al.} 2013). Specifically, metallicities based on oxygen lines are $\sim0.3$ dex lower than those based on nitrogen lines. This difference is attributed to different physical conditions in starforming regions at high redshift, and it is unclear which calibrations, if either, represents the ``true" gas-phase metallicity.

Another uncertainty in our study stems from the fact that, in addition to metallicity, [NII] flux is sensitive to the presence of both AGN and shock excitation. It is unfortunately not possible to separate out AGN contribution to the [NII] flux without high-resolution IFU data ({F{\"o}rster Schreiber} {et~al.} 2011). However, all detected galaxies in our sample have log([NII]/H$\alpha$) $<$ -0.5; the theoretical high-redshift BPT diagram indicates that significant AGN contribution in this regime is unlikely ({Kewley} {et~al.} 2013), though it may exist at a low level. 

We note that the presence of radiative shocks, AGN, and/or different physical conditions at high redshift would imply {\it lower} oxygen abundances than indicated by the measured [NII]/H$\alpha$ ratios. 

It may also be that the stellar oxygen abundances are overestimated. The measurement of oxygen abundance in old unresolved stellar populations is notoriously difficult (see {Conroy} 2013 and references therein). In the {Conroy} \& {van Dokkum} (2012) spectral model, the oxygen abundance is primarily derived from a combination of TiO lines and molecular equilibrium involving CNO. The trend [O/Fe]$\sim$[Mg/Fe] matches that as measured in the Milky Way ({Edvardsson} {et~al.} 1993, though see also {Bensby}, {Feltzing}, \& {Oey} 2013), suggesting that the stellar oxygen abundance measurements are robust.

We also consider the possibility of misidentification of the descendant galaxies. The predicted stellar mass evolution from $z\sim2.3$ to $z=0$ for these galaxies is +0.4 dex. If instead the stellar mass evolution were +0.1/+0.7 dex, this results in negligible change of -0.05/+0.05 dex in the oxygen abundance of their descendants. This is due to the flat relationship between the stellar mass and stellar metallicity ({Panter} {et~al.} 2008). This comparison is thus robust against errors in descendant matching. Additionally, the stellar mass growth at this number density is primarily not through additional star formation, but rather accretion ({van Dokkum} {et~al.} 2010; {Patel} {et~al.} 2013). These accreted galaxies presumably had lower metallicities than the main progenitor galaxy, which means that they would lower the total luminosity-weighted stellar metallicity ({Greene} {et~al.} 2013). Taking this process into account would therefore, if anything, increase the discrepancy in Figure \ref{metals}.

Perhaps the most likely possibility is that selection by stellar mass may still be biased, in the sense that the most metal-rich galaxies do not have detectable emission lines. Only 37.5\% of our sample has detected emission lines; furthermore, the galaxies without detected emission lines are not a random subset of the sample. The last two panels in Figure \ref{ratios} examine the diffe'rent properties of detections and non-detections in our sample. Galaxies with detected emission lines consistently have larger effective radii and higher SFRs than galaxies without detected emission lines.

These properties are known to correlate with the gas-phase metallicity of the galaxies. For example, observations suggest that gas-phase metallicity and SFR are anticorrelated ({Mannucci} {et~al.} 2010). Furthermore, gas-phase metallicity varies by up to 0.2 dex at fixed stellar mass as a function of half-light radius, with larger galaxies having lower metallicities ({Tremonti} {et~al.} 2004; {Ellison} {et~al.} 2008).

Even in galaxies with detected emission lines, some unknown fraction of the star formation may be obscured by dust. Perhaps the detected line flux originates from ``shells" or ``rings" of H$\alpha$, as seen in IFU studies ({F{\"o}rster Schreiber} {et~al.} 2011), while the starforming core remains heavily obscured. UDS-22984 may host such obscured star formation: it has a 24$\mu$m flux indicating a SFR of $\sim270$ M$_{\odot}$/yr and an SED-estimated A$_V$ = 2.6, yet no detected line emission. Since dust correlates with metallicity, this galaxy will likely be more oxygen-rich than galaxies with unobscured star formation. If such heavily obscured star formation is common at high redshift, it remains a possibility that the starforming progenitors of local ellipticals have yet to be detected.

\acknowledgements

We thank the anonymous referee for an outstanding report which substantially improved the paper. Support from HST grant GO-12177 is gratefully acknowledged.



\begin{references}

\reference{} {Adelberger}, K.~L., {Steidel}, C.~C., {Shapley}, A.~E., {et~al.} 2004, \apj,  607, 226

\reference{} {Asplund}, M., {Grevesse}, N., {Sauval}, A.~J., \& {Scott}, P. 2009, \araa, 47,  481

\reference{} {Beers}, T.~C., {Flynn}, K., \& {Gebhardt}, K. 1990, \aj, 100, 32

\reference{} {Bensby}, T., {Feltzing}, S., \& {Oey}, M.~S. 2013, ArXiv e-prints

\reference{} {Brammer}, G.~B., {van Dokkum}, P.~G., \& {Coppi}, P. 2008, \apj, 686, 1503

\reference{} {Brammer}, G.~B., {van Dokkum}, P.~G., {Franx}, M., {et~al.} 2012, \apjs, 200,  13

\reference{} {Bresolin}, F., {Gieren}, W., {Kudritzki}, R.-P., {et~al.} 2009, \apj, 700, 309

\reference{} {Bruzual}, G., \& {Charlot}, S. 2003, \mnras, 344, 1000

\reference{} {Chabrier}, G. 2003, \pasp, 115, 763

\reference{} {Conroy}, C. 2013, ArXiv e-prints

\reference{} {Conroy}, C., {Graves}, G., \& {van Dokkum}, P. 2013, ArXiv e-prints

\reference{} {Conroy}, C., \& {van Dokkum}, P. 2012, \apj, 747, 69

\reference{} {Cullen}, F., {Cirasuolo}, M., {McLure}, R.~J., \& {Dunlop}, J.~S. 2013, ArXiv  e-prints

\reference{} {Dav{\'e}}, R., {Finlator}, K., \& {Oppenheimer}, B.~D. 2012, \mnras, 421, 98

\reference{} {Denicol{\'o}}, G., {Terlevich}, R., \& {Terlevich}, E. 2002, \mnras, 330, 69

\reference{} {Edvardsson}, B., {Andersen}, J., {Gustafsson}, B., {et~al.} 1993, \aap, 275,  101

\reference{} {Ellison}, S.~L., {Patton}, D.~R., {Simard}, L., \& {McConnachie}, A.~W. 2008,  \apjl, 672, L107

\reference{} {Erb}, D.~K., {Shapley}, A.~E., {Pettini}, M., {et~al.} 2006a,  \apj, 644, 813

\reference{} {Erb}, D.~K., {Steidel}, C.~C., {Shapley}, A.~E., {et~al.} 2006b,  \apj, 646, 107

\reference{} {Esteban}, C., {Bresolin}, F., {Peimbert}, M., {et~al.} 2009, \apj, 700, 654

\reference{} {F{\"o}rster Schreiber}, N.~M., {Shapley}, A.~E., {Erb}, D.~K., {et~al.} 2011,  \apj, 731, 65

\reference{} {Gallazzi}, A., {Charlot}, S., {Brinchmann}, J., {White}, S.~D.~M., \&  {Tremonti}, C.~A. 2005, \mnras, 362, 41

\reference{} {Greene}, J.~E., {Murphy}, J.~D., {Graves}, G.~J., {et~al.} 2013, \apj, 776, 64

\reference{} {Horne}, K. 1986, \pasp, 98, 609

\reference{} {Kewley}, L.~J., \& {Dopita}, M.~A. 2002, \apjs, 142, 35

\reference{} {Kewley}, L.~J., \& {Ellison}, S.~L. 2008, \apj, 681, 1183

\reference{} {Kewley}, L.~J., {Maier}, C., {Yabe}, K., {et~al.} 2013, ArXiv e-prints

\reference{} {Kulas}, K.~R., {McLean}, I.~S., {Shapley}, A.~E., {et~al.} 2013, ArXiv  e-prints

\reference{} {Leja}, J., {van Dokkum}, P., \& {Franx}, M. 2013, \apj, 766, 33

\reference{} {Maiolino}, R., {Nagao}, T., {Grazian}, A., {et~al.} 2008, \aap, 488, 463

\reference{} {Mannucci}, F., {Cresci}, G., {Maiolino}, R., {Marconi}, A., \& {Gnerucci}, A.  2010, \mnras, 408, 2115

\reference{} {Marchesini}, D., {van Dokkum}, P.~G., {F{\"o}rster Schreiber}, N.~M., {et~al.}  2009, \apj, 701, 1765

\reference{} {McLean}, I.~S., {Steidel}, C.~C., {Epps}, H.~W., {et~al.} 2012, in Society of  Photo-Optical Instrumentation Engineers (SPIE) Conference Series, Vol. 8446,  Society of Photo-Optical Instrumentation Engineers (SPIE) Conference Series

\reference{} {Newman}, S.~F., {Buschkamp}, P., {Genzel}, R., {et~al.} 2013, ArXiv e-prints

\reference{} {Panter}, B., {Jimenez}, R., {Heavens}, A.~F., \& {Charlot}, S. 2008, \mnras,  391, 1117

\reference{} {Papovich}, C., {Dickinson}, M., \& {Ferguson}, H.~C. 2001, \apj, 559, 620

\reference{} {Patel}, S.~G., {van Dokkum}, P.~G., {Franx}, M., {et~al.} 2013, \apj, 766, 15

\reference{} {Peimbert}, M. 1967, \apj, 150, 825

\reference{} {Pettini}, M., \& {Pagel}, B.~E.~J. 2004, \mnras, 348, L59

\reference{} {Sommariva}, V., {Mannucci}, F., {Cresci}, G., {et~al.} 2012, \aap, 539, A136

\reference{} {Steidel}, C.~C., \& {Hamilton}, D. 1992, \aj, 104, 941

\reference{} {Steidel}, C.~C., {Shapley}, A.~E., {Pettini}, M., {et~al.} 2004, \apj, 604,  534

\reference{} {Stott}, J.~P., {Sobral}, D., {Bower}, R., {et~al.} 2013, ArXiv e-prints

\reference{} {Thomas}, D., {Maraston}, C., {Bender}, R., \& {Mendes de Oliveira}, C. 2005,  \apj, 621, 673

\reference{} {Tremonti}, C.~A., {Heckman}, T.~M., {Kauffmann}, G., {et~al.} 2004, \apj, 613,  898

\reference{} {van der Wel}, A., {Bell}, E.~F., {H{\"a}ussler}, B., {et~al.} 2012, \apjs,  203, 24

\reference{} {van Dokkum}, P.~G., {Whitaker}, K.~E., {Brammer}, G., {et~al.} 2010, \apj,  709, 1018

\reference{} {Whitaker}, K.~E., {van Dokkum}, P.~G., {Brammer}, G., \& {Franx}, M. 2012,  \apjl, 754, L29

\reference{} {Wuyts}, S., {F{\"o}rster Schreiber}, N.~M., {van der Wel}, A., {et~al.} 2011,  \apj, 742, 96

\reference{} {Zahid}, H.~J., {Geller}, M.~J., {Kewley}, L.~J., {et~al.} 2013, \apjl, 771,  L19

\end{references}
\end{document}